\documentclass[aps,prd,twocolumn,superscriptaddress,showpacs,amsmath,amssymb]{revtex4}
	
\usepackage{graphicx,amssymb}
\usepackage{color}
\usepackage{comment}
\usepackage{multirow}
\usepackage{dcolumn}% Align table columns on decimal point
\usepackage{bm}% bold math
\includecomment{printrobustnotes}
\includecomment{printallnotes}
\usepackage{xspace}

\usepackage{amstext}

\begin{document}

%\begin{frontmatter}

\title{Performance of a prototype active veto system using liquid scintillator for a dark matter search experiment}

\author{J.S.~Park}
\affiliation{Center for Underground Physics, Institute for Basic Science (IBS), Daejon 34047, Korea}
\author{P.~Adhikari}
\affiliation {Department of Physics, Sejong University, Seoul 05006, Korea}
\author{G.~Adhikari}
\affiliation {Department of Physics, Sejong University, Seoul 05006, Korea}
\author{S.Y.~Oh}
\affiliation {Department of Physics, Sejong University, Seoul 05006, Korea}
\author{N.Y.~Kim}
\affiliation{Center for Underground Physics, Institute for Basic Science (IBS), Daejon 34047, Korea}
\author{Y.D.~Kim}
\affiliation{Center for Underground Physics, Institute for Basic Science (IBS), Daejon 34047, Korea}
\affiliation {Department of Physics, Sejong University, Seoul 05006, Korea}
\author{C.~Ha}
\affiliation{Center for Underground Physics, Institute for Basic Science (IBS), Daejon 34047, Korea}
\author{K.S.~Park}
\affiliation{Center for Underground Physics, Institute for Basic Science (IBS), Daejon 34047, Korea}
\author{H.S.~Lee}
\affiliation{Center for Underground Physics, Institute for Basic Science (IBS), Daejon 34047, Korea}
\email{hyunsulee@ibs.re.kr}
\author{E.J.~Jeon}
\affiliation{Center for Underground Physics, Institute for Basic Science (IBS), Daejon 34047, Korea}

\begin{abstract}
We report the performance of an active veto system using a liquid scintillator with NaI(Tl) crystals for use in a dark matter search experiment.
When a NaI(Tl) crystal is immersed in the prototype detector, the detector tags 48\% of the internal $^{40}$K background in the 0--10~keV energy region.
We also determined the tagging efficiency for events at 6--20~keV as 26.5 $\pm$ 1.7\% of the total events, which corresponds to 0.76 $\pm$ 0.04~events/keV/kg/day.
According to a simulation, approximately 60\% of the background events from U, Th, and K radioisotopes in photomultiplier tubes are tagged at energies of 0--10~keV.
Full shielding with a 40-cm-thick liquid scintillator can increase the tagging efficiency for both the internal $^{40}$K and external background to approximately 80\%. 

\end{abstract}

%\begin{keyword}
%Dark matter, WIMP, low-background techniques, background veto, liquid scintillator, radioactivity
%\end{keyword}
%% PACS codes here, in the form: \PACS code \sep code

%\end{frontmatter}
\maketitle

%% \linenumbers

%% main text
\section{Introduction}
\label{introduction}

Numerous astronomical observations have indicated that most of
the matter in the universe is invisible, exotic, and nonrelativistic dark
matter~\cite{dclowe,Komatsu:2010fb,Ade:2013zuv,gbertone}. However, despite considerable experimental effort, the nature of the dark matter remains unknown.
The weakly interacting massive particle~(WIMP) is a candidate that conforms to the stringent standards for particle
dark matter, and it is supported by astronomical observations as well as particle physics theories that
extend the standard model~\cite{lee77,jungman96}. Numerous experiments have searched for WIMPs
by directly detecting nuclei recoiling from WIMP--nucleus interactions~\cite{dmsummary,tmarr}.

It is essential to reduce the background level
as much as possible for dark matter searches because of the extremely low cross-section of the WIMP--nucleon interaction. The dominant background contributions produced by external sources, such as decaying radioisotopes in surrounding materials and cosmogenic muons,
can be reduced by shielding the detector in a deep underground laboratory using heavy materials consisting of lead, copper, and polyethylene layers. However, remnants of this contribution, as well as the background contribution from materials close to the detector,
still exist. In addition to passive shielding, it is beneficial to have an active veto system that can reduce the background level by tagging neutron and $\gamma$ events~\cite{lsveto1,lsveto2,lsveto3} from external sources. An active veto system can also reduce the level of the internal background by tagging the escaping $\gamma$-rays.

The Korea Invisible Mass Search (KIMS) Collaboration is developing
ultralow-background NaI(Tl) crystal detectors~\cite{kims_nai1,kims_nai2,kims_nai3} to verify an observation of an annual modulation in the detection rate
of an array of NaI(Tl) crystals~\cite{bernabei13} that can be interpreted as the result of interaction with WIMPs.
Even though the internal radioisotopes of the NaI(Tl) crystals were identified and might be further reduced by future development, the background contributions originating
from detector components, such as photomultiplier tubes~(PMTs), encapsulating copper cases, and shielding materials, are not perfectly rejected. 
We consider a new shielding structure including an active veto system that surrounds the NaI(Tl) crystals with a liquid scintillator~(LS) in order to tag the internal $^{40}$K background as well as 
reduce the external background contributions.
We build a prototype active veto detector to verify the performance of the full shielding before finishing its final design. 
In this article, we describe the prototype active veto detector from construction to performance. In addition, a shielding design for a future NaI(Tl) experiment is presented.

\section{Prototype liquid scintillator veto system}

\begin{figure*}[!htb]
\begin{center}
		\begin{tabular}{ccc}
\includegraphics[width=0.44\textwidth]{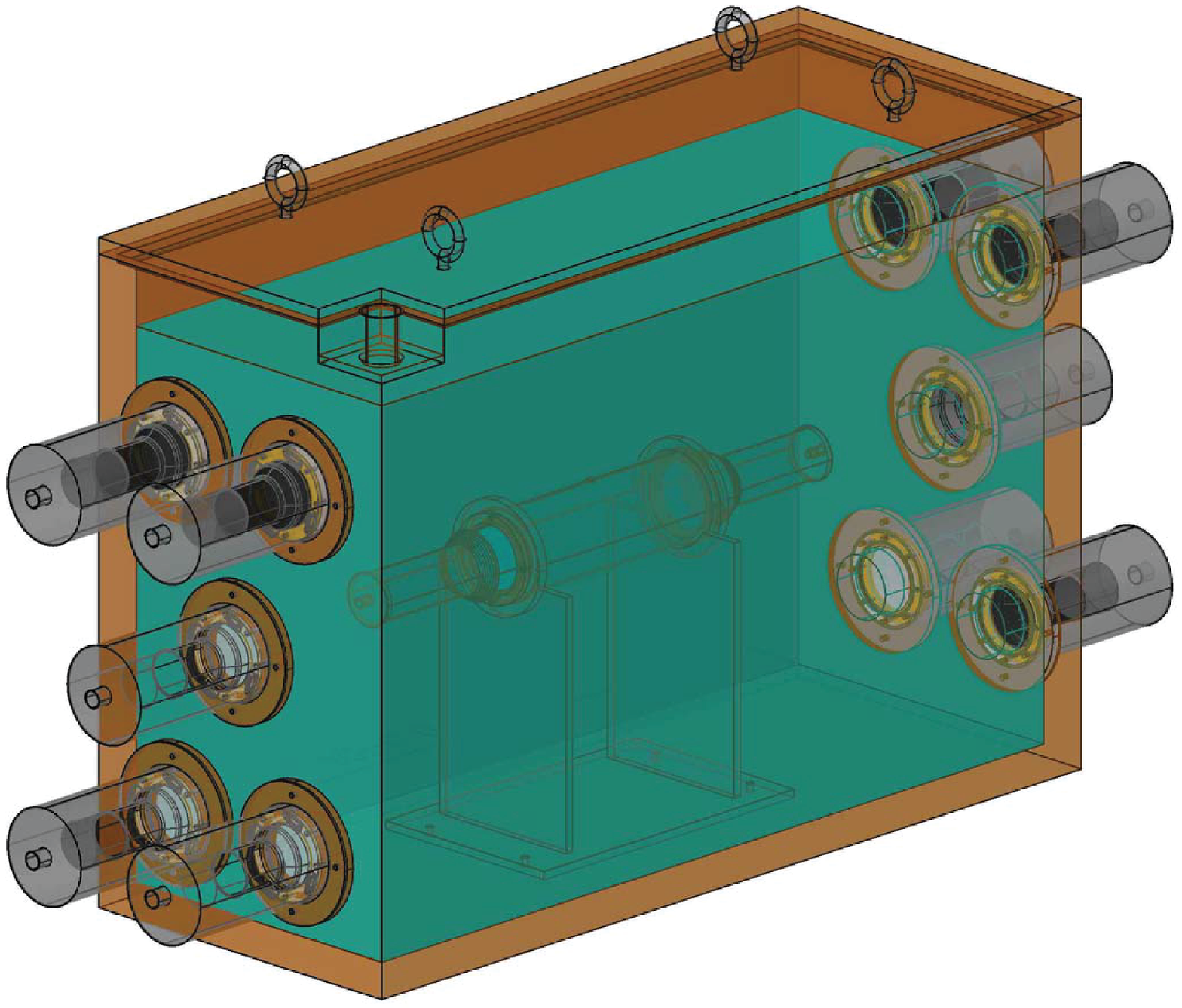} &
\includegraphics[width=0.283\textwidth]{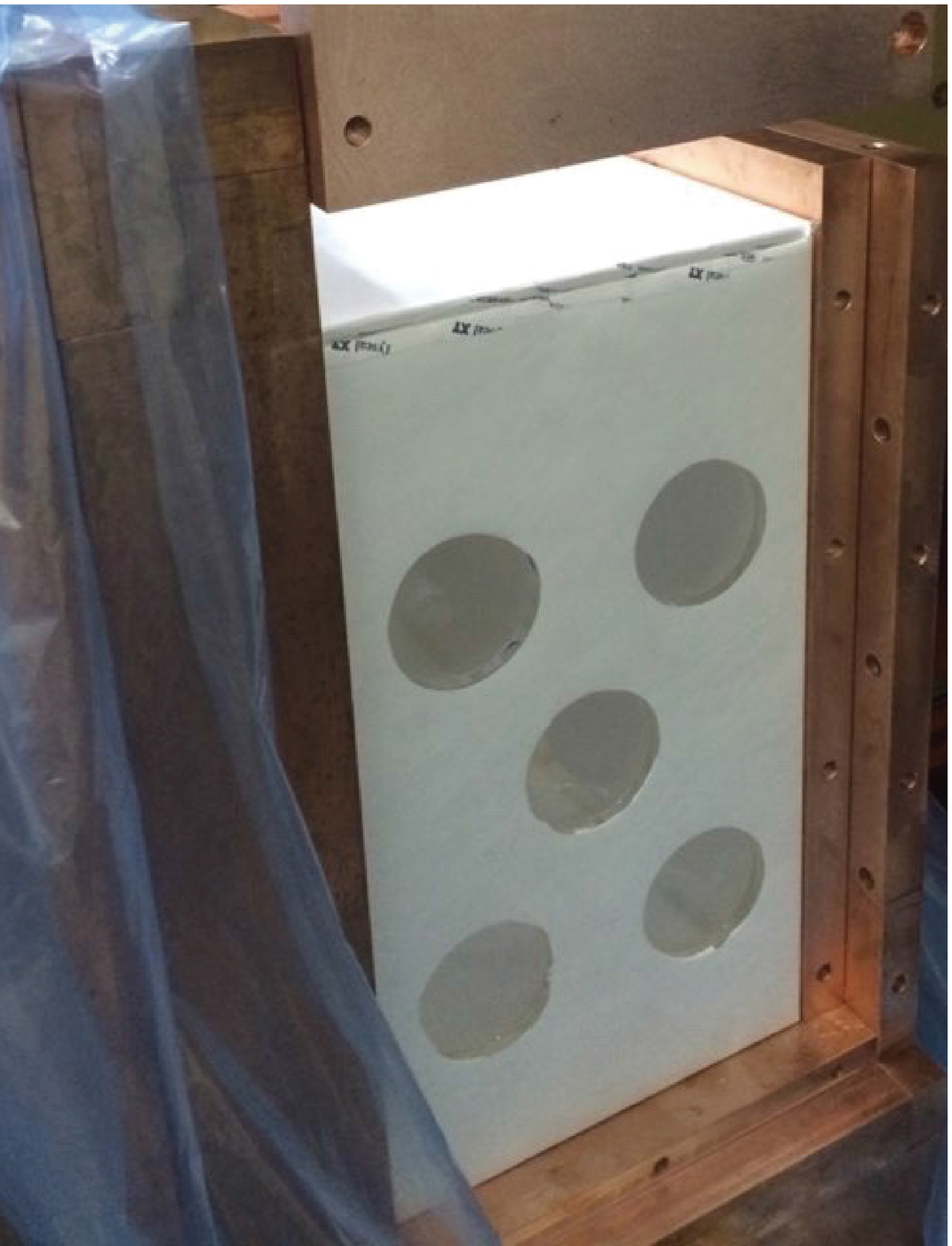} &
\includegraphics[width=0.23\textwidth]{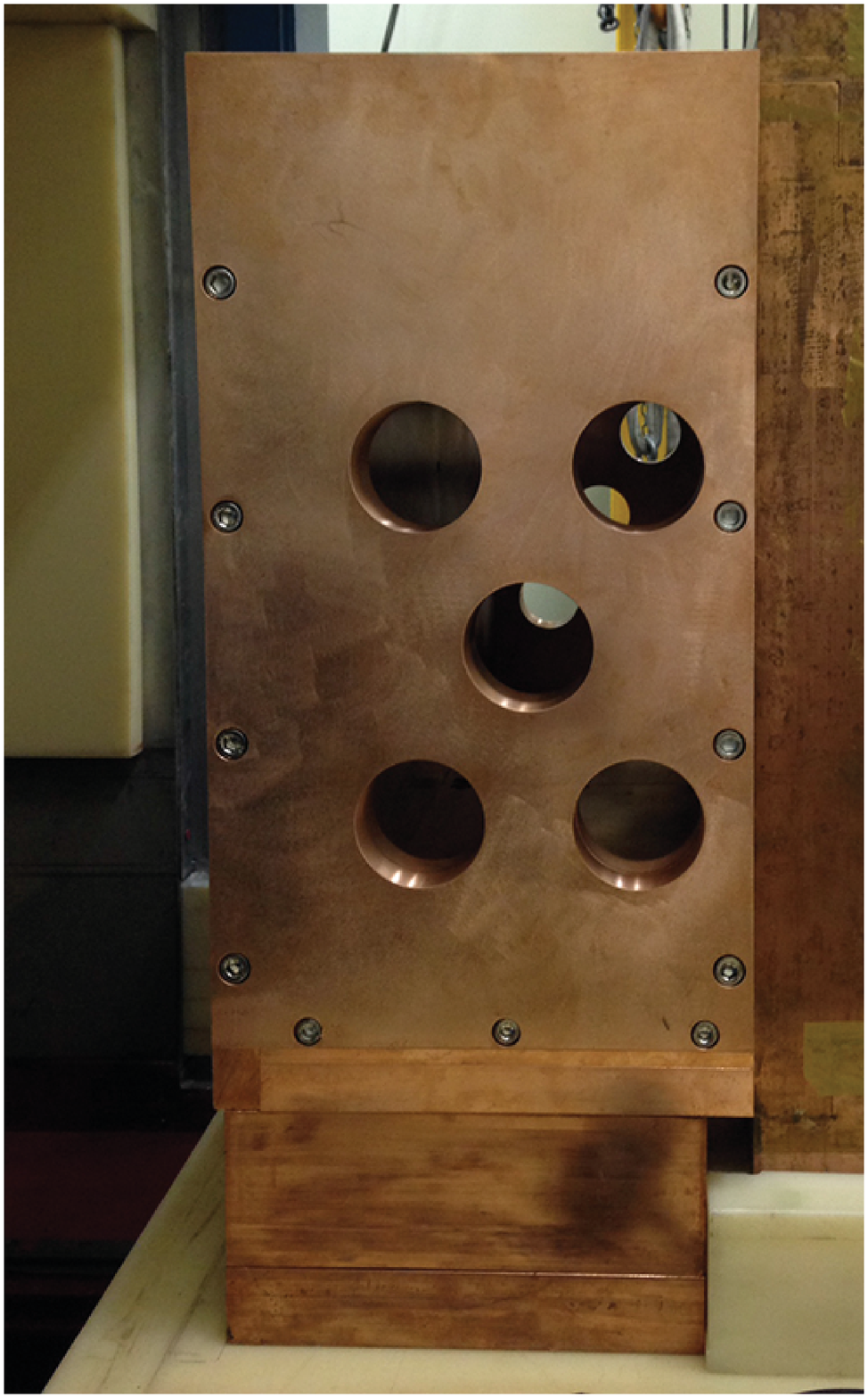} \\
(a)  & (b) & (c) \\
\end{tabular}
\end{center}
\caption{ (a) Design of prototype LS veto detector and installation structure consisting of (b) acrylic box and (c) copper box.
  }
\label{experimentalsetup}
\end{figure*}

We produce a linear alkyl benzene (LAB)-based LS~\cite{labls,reno_jspark} consisting of 3~g of 2,5-diphenyloxazole~(PPO)/$\ell$ and 30~mg of 1,4-bis[2-methylstyryl]benzene~(bis-MSB)/$\ell$
for a high light yield and long attenuation length. 
We use a water extraction method~\cite{water_extraction} and nitrogen purging to remove possible contamination from natural radioisotopes in the LS.
The measured contamination levels from $^{238}$U and $^{232}$Th in the LS are less than 7 and 4~ppt, respectively. These contaminants provide negligible background to the NaI(Tl) crystal. 

\begin{figure*}
\begin{center}
\includegraphics[width=0.9\textwidth]{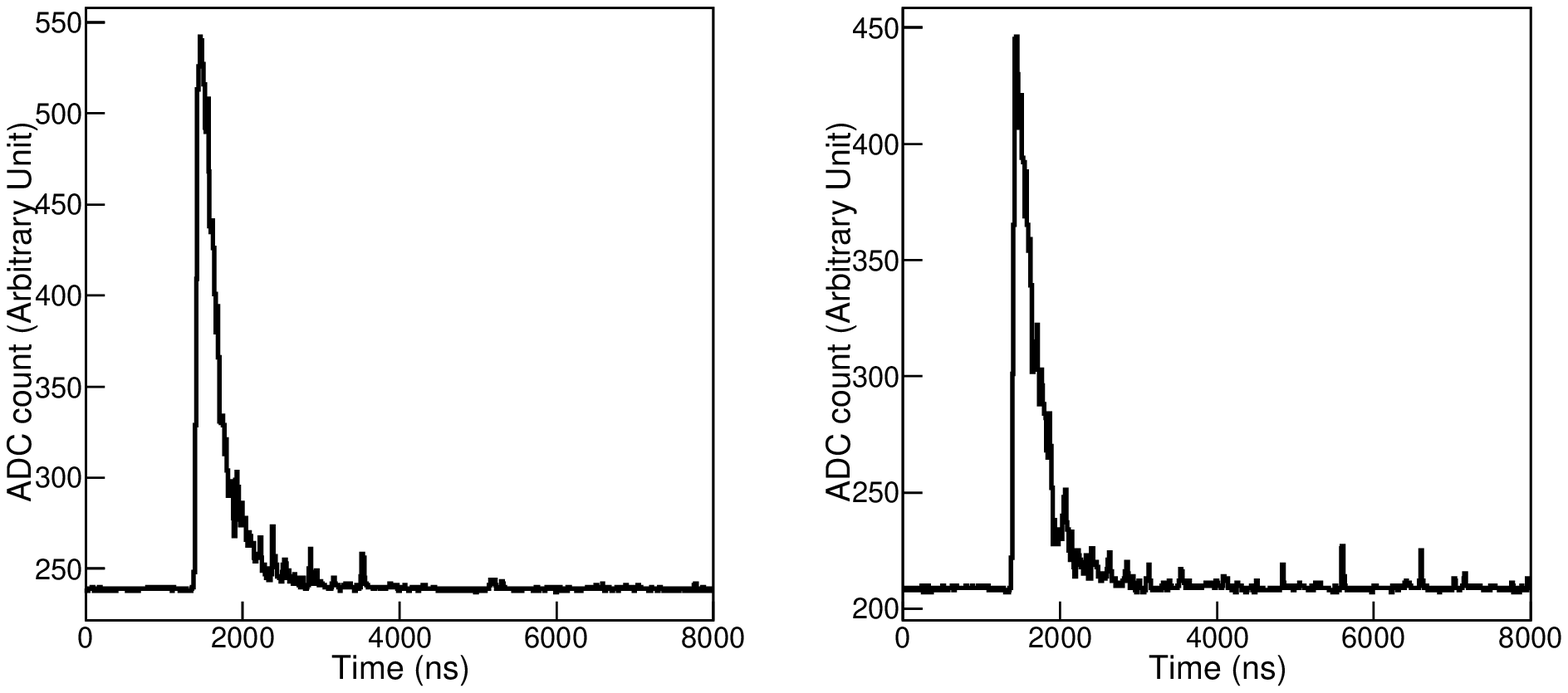} \\
NaI crystal\\
\includegraphics[width=0.9\textwidth]{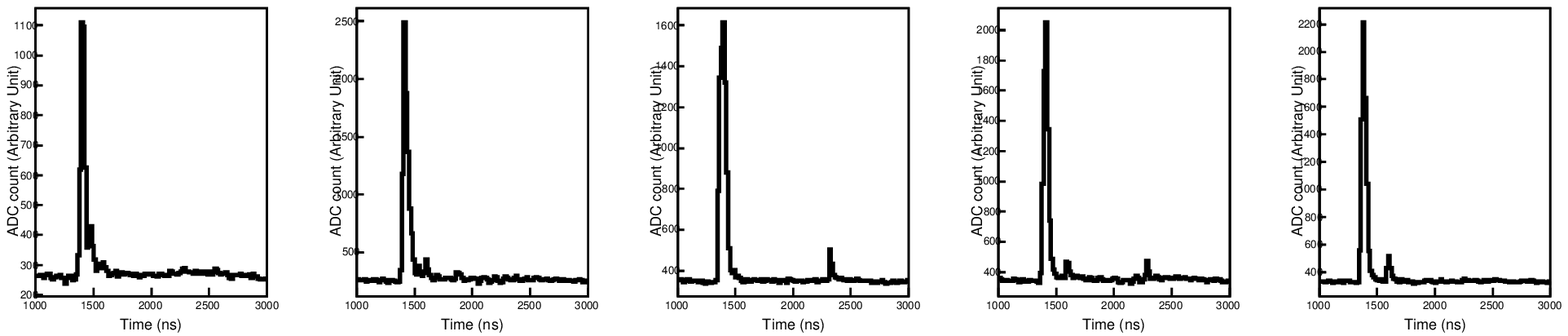} \\
LS veto (left side)\\
\includegraphics[width=0.9\textwidth]{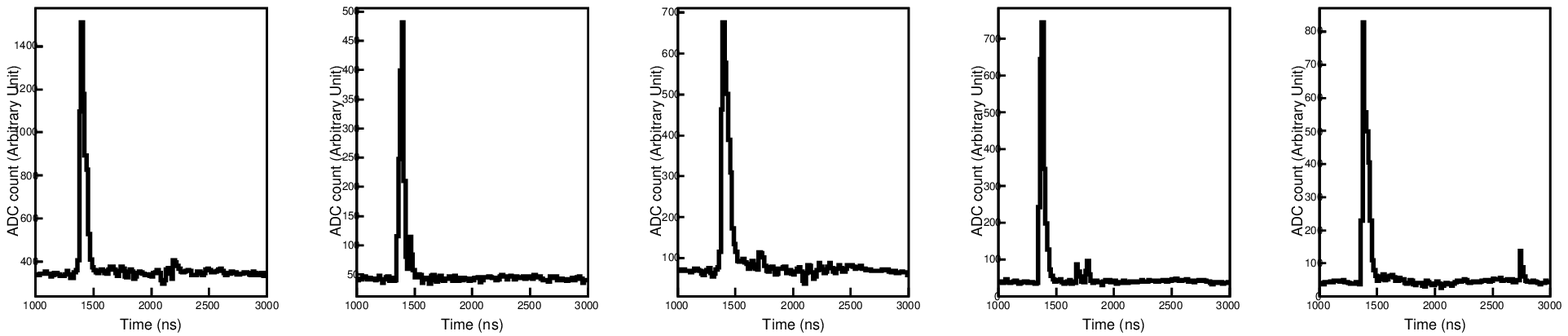} \\
LS veto (right side)\\
\caption[Typical event]{Pulse shapes from PMTs for typical coincident events of the NaI crystal and LS veto detector with measured energies of 37 and 850~keV, respectively.
}
\label{ref:fig.lsveto.typical}
\end{center}
\end{figure*}

To take advantage of a low-background facility, the prototype detector is designed to fit in an existing shield for the KIMS-CsI experiment at Yangyang Underground Laboratory (Y2L)~\cite{hslee06, hsleenim, sckim12}.
As shown in Fig.~\ref{experimentalsetup}, the prototype detector is relatively narrow because of the limited space in the shielding.
The maximum available space surrounded by the layers of 5~cm polyethylene and 10~cm copper accommodates a 110~cm (L) $\times$ 44~cm (H) $\times$ 25~cm (W) rectangular acrylic box, including an additional 6~cm of copper on the outside.
The acrylic structure can contain 140~$\ell$ of the LAB-based LS. A separate acrylic support is installed to hold a NaI(Tl) crystal in the middle of the box. The crystal is wrapped in a Teflon sheet and encapsulated in oxygen-free high-conductivity (OFHC) copper with two synthetic quartz windows that allow coupling to the PMTs. The PMTs are encased in another OFHC copper housing to decouple light signals between the crystal and the LS. % as well as to avoid direct contact of the PMTs to LS. 

The acrylic box is wrapped in Tyvek reflective films~[see Fig.~\ref{experimentalsetup} (b)] to increase the light collection efficiency of the LS. Scintillation signals produced by the LS are measured by ten 3~inch Hamamatsu low-radioactivity and high-quantum-efficiency PMTs (R12669SEL).
The PMT signals are amplified by a factor of 30; the amplified signals are digitized by 64~MHz flash analog-to-digital converters (FADCs). The total recorded time window for an event is 8~$\mu$s.
The energy scale of the LS veto detector is calibrated using the Compton edges of $\gamma$-ray events from a $^{60}$Co~(1173~keV and 1333~keV $\gamma$ lines) source and extrapolated to low energy.
The energy calibration is checked with $^{57}$Co~(122.1~keV $\gamma$ line) and $^{137}$Cs~(661~keV $\gamma$ line) sources, which show good agreement. 

After the structure is installed, an already tested NaI(Tl) crystal (NaI-002)~\cite{kims_nai1} is placed in the center, and the LS is fitted into the acrylic box.
NaI-002 was grown from selected ultrapure NaI powder by the Alpha Spectra Company. The crystal~(9.2~kg mass) has a cylindrical shape with a diameter of 4.2~inch and a length of 11~inch. 
Two R12669SEL PMTs are attached at each end of the crystal. The PMT signals from the crystal are split and amplified by factors of 30 and 2. The amplified signals are digitized by 400 and 64~MHz FADCs with readout time windows of 10 and 8~$\mu$s. A trigger condition for the NaI(Tl) crystal is the presence of one or more photoelectrons in each PMT within a 200~ns time window.
If the NaI(Tl) crystal is triggered, all the signals from both the NaI(Tl) crystal and the LS veto detector are recorded.
The typical event shapes, which show signals from both NaI and LS, are shown in Fig.~\ref{ref:fig.lsveto.typical}.
Low-energy calibration of the NaI(Tl) crystal was done using the 59.54~keV $\gamma$-line from a $^{241}$Am source. An extrapolation of the 59.54~keV line to low energy was checked using the 3~keV X-ray line from internal $^{40}$K, which agreed within 5\%. 

\begin{figure*}
\begin{center}
\begin{tabular}{ccc}
\includegraphics[width=0.33\textwidth]{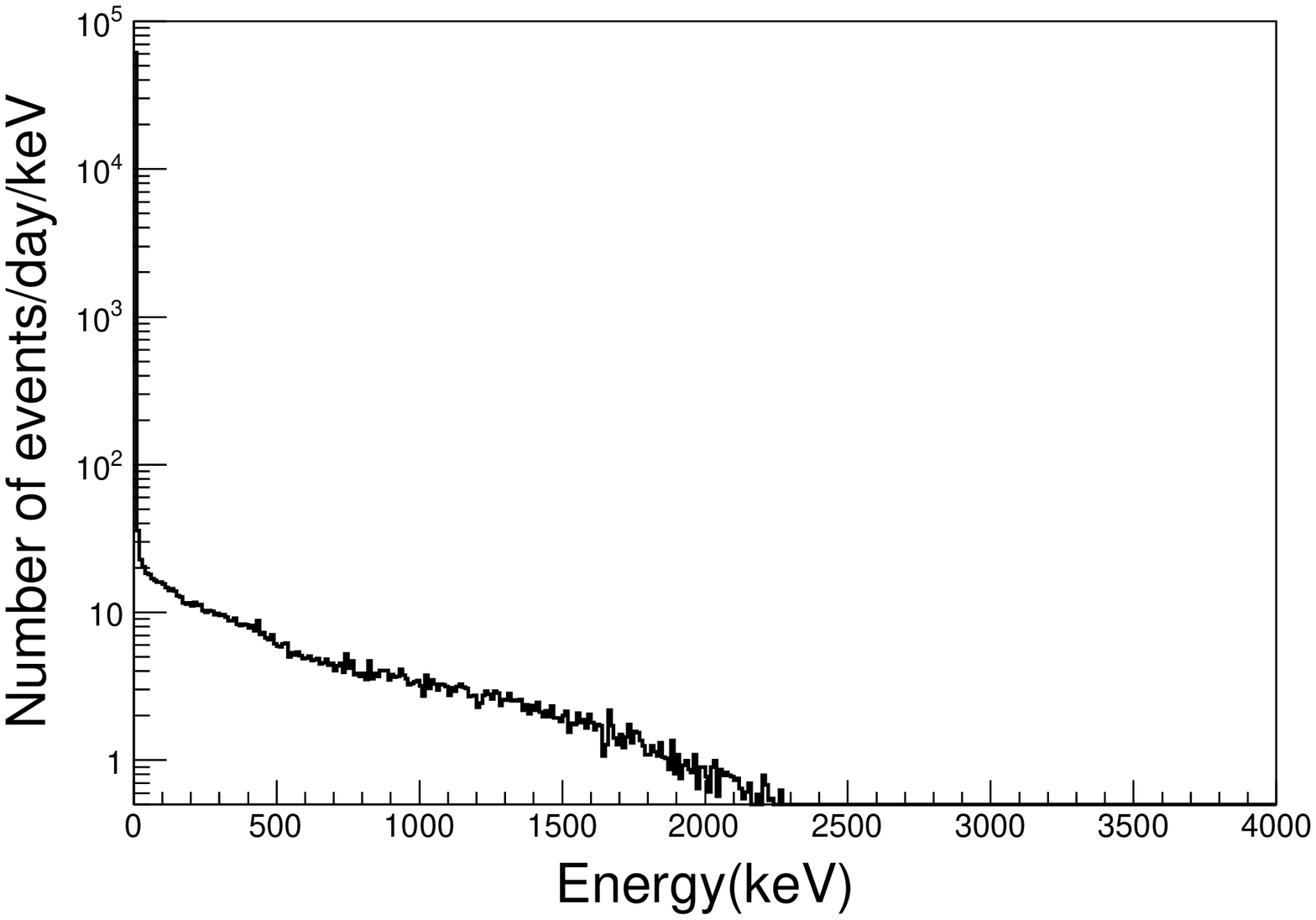} &
\includegraphics[width=0.33\textwidth]{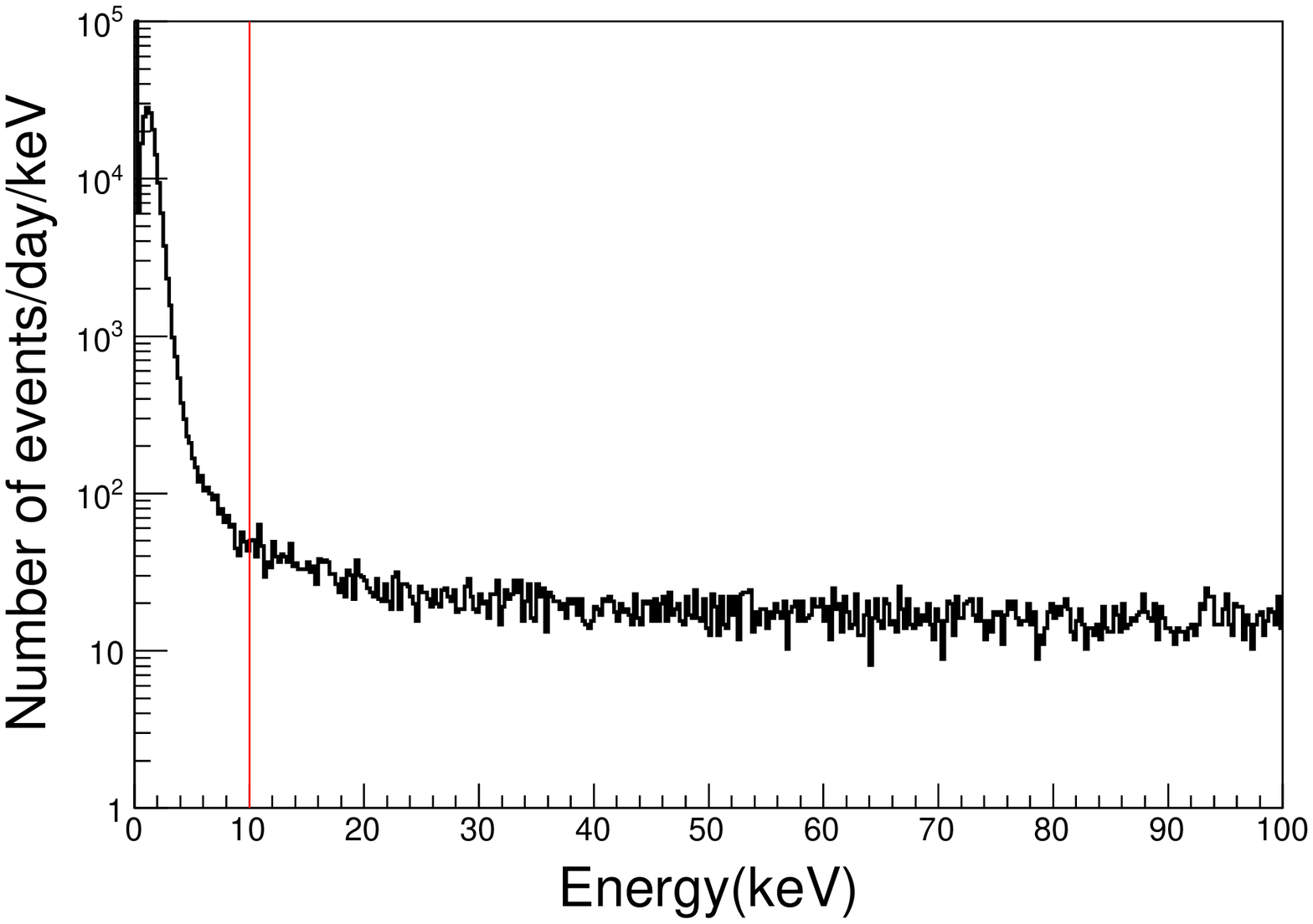} &
\includegraphics[width=0.33\textwidth]{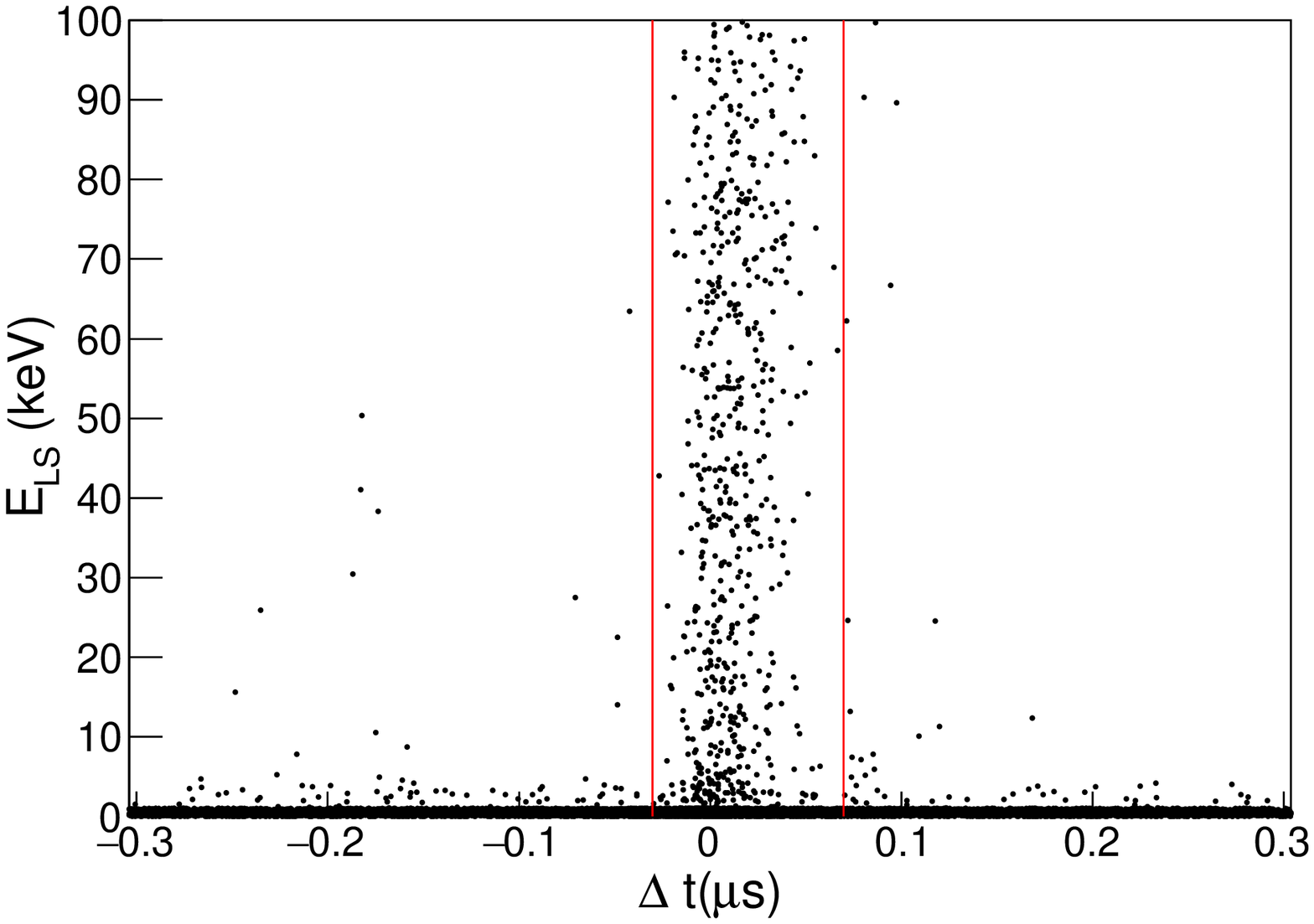} \\
(a) & (b) & (c) \\
\end{tabular}
\caption[Veto detector energy]{Energy spectra of LS veto detector. (a) Energy spectrum for $\gamma$ background region of 0--4~MeV. (b) Zoomed energy spectrum in low-energy region (0--100~keV) with 10~keV energy threshold~(red line) for coincidence condition. (c) Energy of LS veto detector versus time difference between LS veto signal and NaI(Tl) signal~($\Delta t$) for coincidence condition~(red line). 
}
\label{ref:fig.lsveto.vetosignal}
\end{center}
\end{figure*}

Figure~\ref{ref:fig.lsveto.vetosignal} (a) and (b) show the energy spectra of the LS veto detector in various
energy ranges. Figure~\ref{ref:fig.lsveto.vetosignal}~(c) shows the energy of the LS veto detector as a function of the time difference between the LS veto signal and the NaI(Tl) signal~($\Delta t$). 
With the requirement of 10~keV of energy from the LS, one can remove dominant random events so that the crystal-related background coinciding with the LS veto detector can be efficiently tagged. 
In addition, we require $\Delta t$ to be 
$-30~\mathrm{ns}<\Delta t<70~\mathrm{ns}$. The event rate of the LS veto detector passing the criteria is only 0.12~Hz, which makes the veto-induced dead time negligible. We can calculate a fake event rate of 0.1\% in the coincident event region from the events occurring outside the coincident region~($-1000~\mathrm{ns}<\Delta t < -30~\mathrm{ns}$ or $70~\mathrm{ns}<\Delta t < 1000~\mathrm{ns}$).

\begin{figure*}
\begin{center}
\begin{tabular}{cc}
\includegraphics[width=0.49\textwidth]{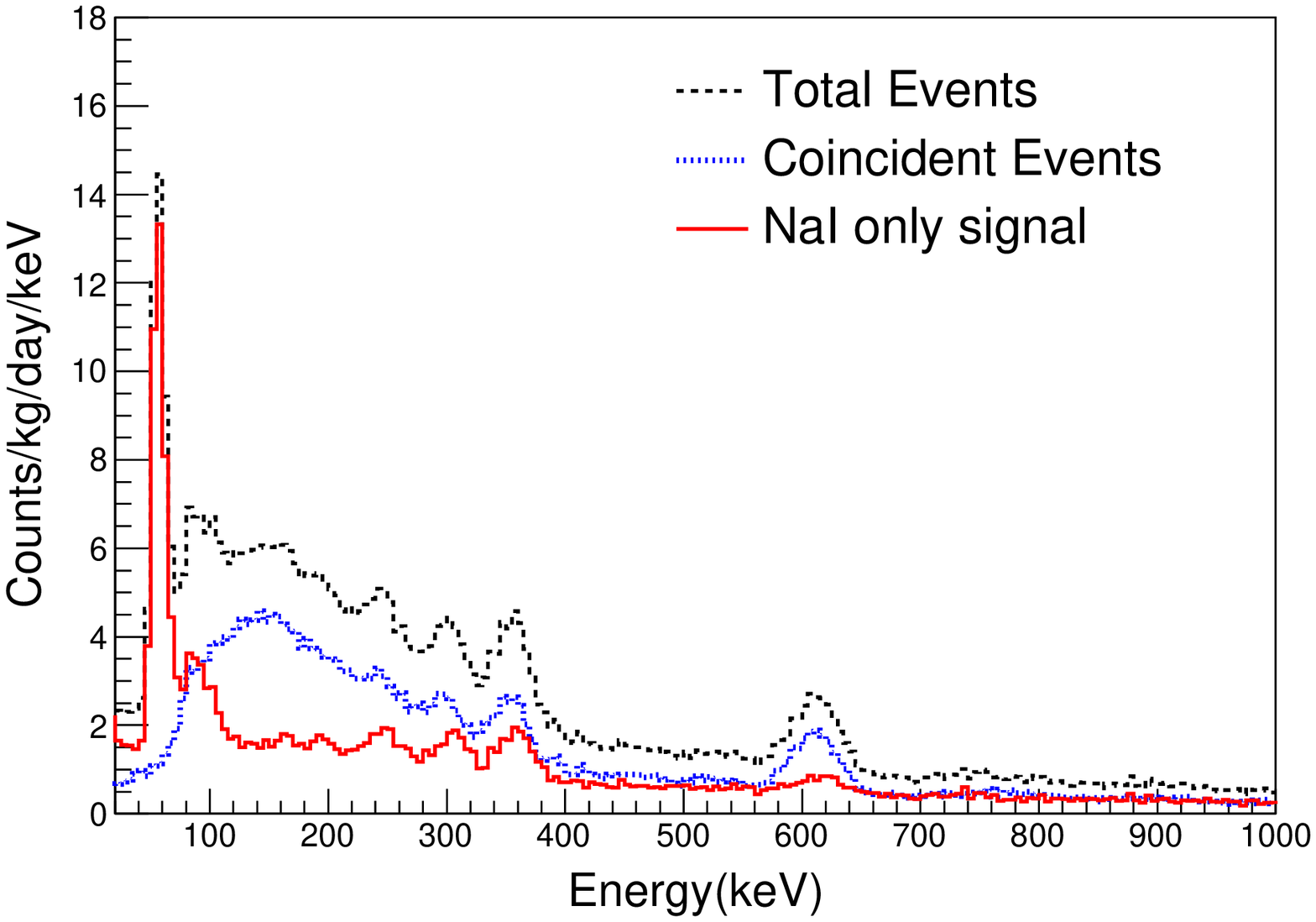}&
\includegraphics[width=0.49\textwidth]{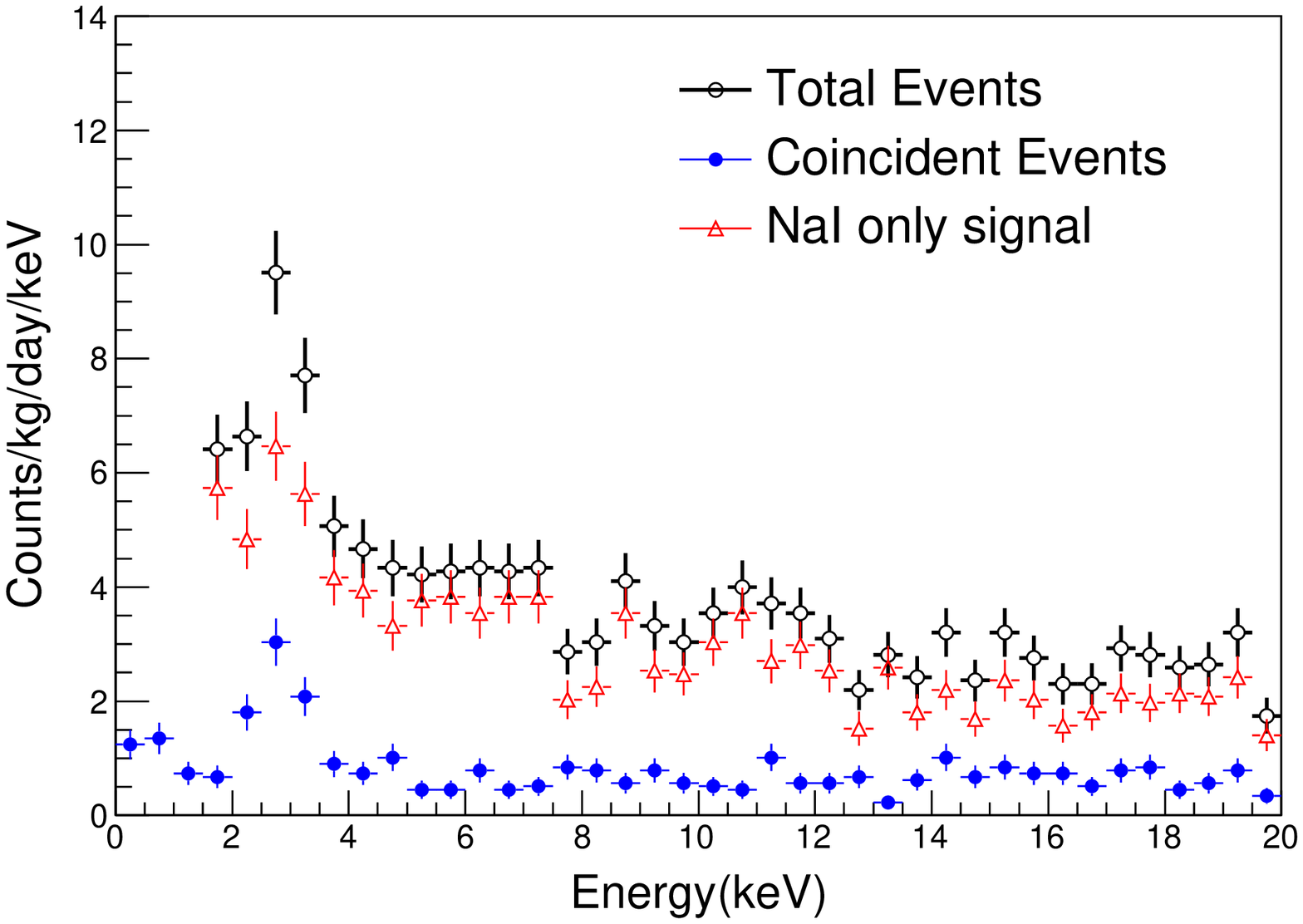}\\
(a) High Energy & (b) Low Energy \\
\end{tabular}
\caption{Background energy spectra of the NaI(Tl) crystal in the LS veto system. Vetoed events (blue symbols) have hits on the LS veto detector with energies greater than 10~keV, whereas hits on NaI only~(red) have no such hits on the LS veto detector. The total hits~(black) represents all events (the sum of events in both categories).
}
\label{ref:fig.lsveto.nai}
\end{center}
\end{figure*}

We count the event rate of the NaI(Tl) detector with and without the LS veto requirement, as shown in Fig.~\ref{ref:fig.lsveto.nai}. The detector realizes a significant reduction in the background rate by requiring that no signal appear in the LS veto detector. The tagging efficiency is 26.5 $\pm$ 1.7\% in the 6--20~keV energy region and 63 $\pm$ 1\% in the 100--1500~keV energy region. The tagged event rate between 6 and 20~keV is 0.76 $\pm$ 0.04~events/keV/kg/day (dru).

We simulate various background sources with the Geant4~\cite{geant4} simulation package to obtain the veto efficiency of the prototype. 
We first consider the internal $^{40}$K decay, which is one of the strongest backgrounds for dark matter searches with NaI(Tl) crystals~\cite{anais,dmice,picolon,sabre}. This decay generates an X-ray at approximately 3~keV with a 1460~keV $\gamma$-ray. If the accompanying 1460~keV $\gamma$-ray escapes from the crystal, the event consists of a single 3~keV hit, which mimics a dark matter signal. However, if we tag the escaping 1460~keV $\gamma$-ray with the LS veto detector, we can effectively reduce the low-energy contribution of the 3~keV X-ray in the NaI(Tl) crystal~\cite{anais1}. The veto efficiency for energies between 0 and 10~keV is about 48\% from the internal $^{40}$K simulation. 
We estimate the K contamination of the NaI(Tl) crystal using the events vetoed by the LS veto detector to verify our simulation. 
%Using low energy measured spectrum of the NaI(Tl) crystal~(blue symbols in Fig.~\ref{ref:fig.lsveto.nai}~(b)), 
The filled (blue) circles in Fig.~\ref{ref:fig.lsveto.nai}~(b) show the low-energy spectrum of the NaI(Tl) crystal that is tagged by the LS veto detector. In this figure, the tagged 3~keV X-ray events can be extracted by Gaussian fitting assuming a flat background. The measured rate is 2.10~$\pm$~0.32 counts/kg/day, which is found to correspond to 43.7~$\pm$~6.4~ppb K contamination by a comparison with the simulation. This value is consistent with a previous measurement~(49.3~$\pm$~2.4~ppb) of the same crystal in a different veto system in which neighboring CsI(Tl) crystals are used to tag the $^{40}$K events~\cite{kims_nai1}.

%\begin{figure}
%\begin{center}
%\includegraphics[width=0.7\textwidth]{./multiple_sim.eps}
%\caption{Tagging rate of 3~keV X-rays caused by internal $^{40}$K decay calculated using the vetoed events. A comparison with a Geant4 simulation yields a K contamination of 43.7 $\pm$ 6.4~ppb.
%}
%\label{ref:fig.lsveto.k40}
%\end{center}
%\end{figure}

Other than the internal background of the NaI(Tl) crystal, one of the main background contributions arises from the natural radioisotopes, $^{238}$U, $^{232}$Th, and $^{40}$K, in the PMTs.
We simulate these isotopes from the photocathode as well as the glass envelope and estimate the veto efficiencies in the 0--10~keV measured energy range, which are
66\%, 56\%, and 63\% for $^{238}$U, $^{232}$Th, and $^{40}$K, respectively.
Other background contributions from outside the LS veto system are tagged at a level that is at least similar to that of the PMTs. 

\section{Design of a new LS veto system for a 200~kg NaI experiment}

\begin{figure}
		\begin{center}
\includegraphics[width=0.8\columnwidth]{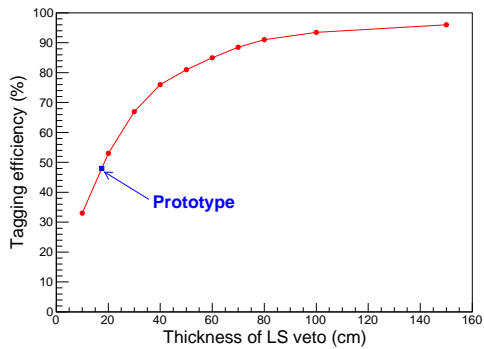} 
\caption[Veto efficiency curve]{Veto efficiency of internal $^{40}$K as a function of the thickness of the NaI(Tl) crystal and the LS container for energies between 0 and 10~keV obtained from a simple geometric simulation. Square symbol represents the veto efficiency of the prototype detector from the simulation. 
}
\label{ref:fig.veto.eff.curve}
		\end{center}
\end{figure}

\begin{figure}
\begin{center}
\includegraphics[width=0.8\columnwidth]{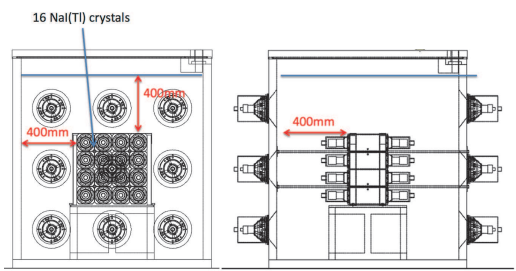}
\caption{Drawing of an LS veto system for a 200~kg NaI(Tl) crystal array. Nine 5~in. Hamamatsu PMTs will be attached to each end of
the Cu container.}
\label{ref:fig.MainCuBox0}
\end{center}
\end{figure}

The veto efficiency of the prototype detector is limited by the narrow thickness of the container necessitated by the small available space in the existing shielding at Y2L. 
We plan to construct a dedicated shielding structure for the 200~kg array experiment using NaI(Tl) crystals~\cite{kims_nai3}. In this shielding structure,
we can consider a much bigger container for the LS veto system to increase the veto efficiency.
The optimization is based on the simulation of the internal $^{40}$K background. 
We consider a simple cubic structure of the LS container in which one NaI(Tl) crystal~(same geometry as NaI-002) is immersed in the center. The thicknesses of the NaI(Tl) crystal ends and the LS container wall are the same and vary
from 10 to 150~cm. Figure~\ref{ref:fig.veto.eff.curve} shows the veto efficiency as a function of the LS thickness. 
We consider an LS with a minimum thickness of 40~cm for the final design by considering the size of the shielding as well as the veto efficiency. 
The veto efficiency measured by the prototype detector is consistent with the simulated efficiency for a 17-cm-thick container, as shown in Fig.~\ref{ref:fig.veto.eff.curve}.

An LS veto detector 40~cm thick can tag approximately 75\% of the internal $^{40}$K events at 0--10~keV. The untagged 25\% of events are due to 1460~keV $\gamma$-rays escaping without any heat in the LS veto detector. 
The background contribution from the internal $^{40}$K with the new veto design can be effectively reduced by a factor of two because the 25\% of untagged events represents a reduction of approximately two times compared to the 52\% of untagged events with the prototype detector.

We finalized the design of the new LS veto system, as shown in Fig.~\ref{ref:fig.MainCuBox0}, considering the discussion above. The minimum thickness of the LS from each end of the crystal is 40~cm. 
The average tagging efficiency of the LS veto system and the other crystals for the internal $^{40}$K is estimated to be 81\%. The veto efficiencies for the PMTs are predicted to be approximately 80\% for U, Th, and K. 

\section{Conclusion}
We constructed a prototype LS veto detector to determine the feasibility of tagging the background of a NaI(Tl) crystal detector for a dark matter search experiment.
The tagged background rate of the NaI(Tl) crystal is 0.76 $\pm$ 0.04~dru at 6--20 keV. 
The tagging efficiency of the 3~keV X-rays from internal $^{40}$K is about 48\% from the simulation. The measured event rates of 3~keV X-rays with the required LS veto signal are in good agreement with a previous measurement. 
%The internal $^{40}$K rejection is about 48\%, which is consistent with the observed data. 
A new shield with an appropriate LS thickness will increase this rate to approximately 80\%.

\section*{Acknowledgments}
We thank the Korea Hydro and Nuclear Power (KHNP) Company for
providing the underground laboratory space at Yangyang. This research
was funded by Grant No. IBS-R016-A1 and was supported by the Basic
Science Research Program through the National Research Foundation of Korea
(NRF) funded by the Ministry of Education~(NRF-2011-35B-C00007).

\end{document}